\documentclass[twocolumn]{revtex4}
\usepackage{graphicx}
\usepackage{scrtime}
\usepackage{latexsym}
\usepackage{epsfig}
\usepackage{fancyhdr}
\usepackage[hang]{subfigure}
\usepackage{supertabular}
\usepackage{wrapfig}
\usepackage{amssymb}

\begin{document}

\title{Four-point measurements of n- and p-type two-dimensional systems fabricated with cleaved-edge overgrowth}

\author{M. Grayson}
\author{S. F. Roth}
\author{Y. Xiang}
\author{F. Fischer}
\author{D. Schuh}
\author{M. Bichler}

\affiliation{Walter Schottky Institut, Technische Universit\"at M\"unchen, 85748, Garching, Germany}

\begin{abstract}
\noindent We demonstrate a contact design that allows four-terminal magnetotransport measurements of cleaved-edge overgrown two-dimensional electron and hole systems. By lithographically patterning and etching a bulk-doped surface layer, finger-shaped leads are fabricated, which contact the two-dimensional systems on the cleave facet.  Both $n$- and $p$-type two-dimensional systems are demonstrated at the cleaved edge, using Si as either donor or acceptor, dependent on the growth conditions.  Four-point measurements of both gated and modulation-doped samples yield fractional quantum Hall features for both $n$- and $p$-type, with several higher-order fractions evident in $n$-type modulation-doped samples. 
\end{abstract}

%\pacs{}

\maketitle
Cleaved-edge overgrowth (CEO) was introduced with a two-point measurement of a two-dimensional (2D) electron system grown on an in-situ cleaved facet \cite{pfeiffer:1697}.  Since then, this technique has been adapted to create various low-dimensional $n$-type transport structures \cite{grayson:233,yacoby:77,kang:59,deutschmann:2175,auslaender:5556,huber:016805}.  $p$-type devices with the same structure as the $n$-type promise to reveal different physics due to the large hole mass and strong spin-orbit coupling, and recently the first $p$-type CEO structure was demonstrated in the form of a quantum wire \cite{pfeiffer:073111}.  To date, however, transport measurements of a CEO-grown 2D hole system have not been reported.  In the effort to optimize CEO growth for both $p$- and $n$-type structures, it would also be helpful if 4-point measurements could be performed to characterize the overgrown interface.

In this Letter, we present a contact design for 4-point transport characterization on the cleave facet, which we demonstrate for both $n$- and $p$-type samples.  Patterned contact fingers are made of either $n^+$-GaAs or $p^+$-GaAs, and intersect the 2D system grown on the cleave plane.  CEO mobilities reach $4 \times 10^{6} ~\mathrm{cm^2/Vs}$ and manifest higher order fractions in the quantum Hall effect.  Gated samples are grown by CEO and tune the 2D density continuously.  We also demonstrate 4-point measurements of a 2D hole system with $p$-type CEO.  

\begin{figure}[!ht]
\center
\includegraphics[width=\linewidth ,keepaspectratio]{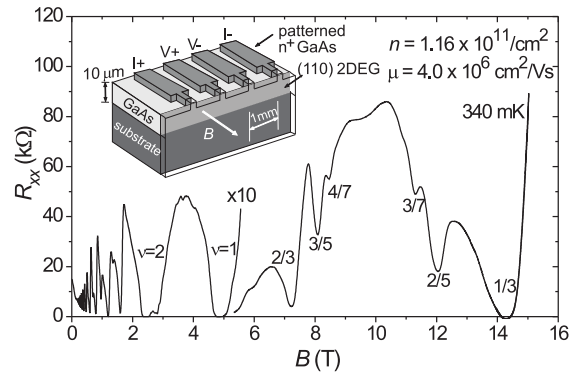}
\caption{Longitudinal resistance $ R_{xx}$ of a modulation doped 2D electron system measured in a four-terminal geometry. Inset: lithographically defined $ n^+$-fingers conduct to the 2D system as current and voltage contacts. Minima at fractional filling factors $\nu$ prove the high quality of the sample.}
\label{nRXX}
\end{figure}

First, we consider $n$-type samples \cite{sample:nfinger}. A semi-insulating (001) substrate is overgrown with a buffer layer and 10 $\rm \mu m$ of undoped GaAs.  This is topped by $\rm 1\: \mu m$ of $ n^+$ Si-doped GaAs which is photolithographically patterned and etched, leaving a row of $n^+$-fingers (see inset, Fig.~\ref{nRXX}). Etching structures more than 1 $\mu$m deep was observed to result in bad cleaves.  The distance between the individual fingers is 1 mm, and the width of the finger at the cleave plane is of order 10 $\mu$m.
After reinserting the sample into the growth chamber, it is cleaved in-situ and the freshly exposed perpendicular (110) cleavage plane is overgrown at optimized CEO growth conditions with a substrate pyrometer temperature $T_\mathrm{pyr} = 490^\mathrm{o}$C and an As$_4$ beam flux of $P_\mathrm{As_4} = 3.3 \times 10^{-5}$ mbar.  A 420~nm $\rm Al_{0.3}Ga_{0.7}As$ barrier is grown, capped with $\rm 20\:nm$ GaAs, and the barrier contains Si $\delta$-doping at a spacer distance $\rm 120\:nm$ from the cleave interface. Instead of the cap layer, $n^+$-doped GaAs can complete the modulation-doped heterostructure and function as a sidegate. A 2D electron system forms at the GaAs/AlGaAs cleave interface and is contacted by the $ n^+$-fingers, which terminate at the cleave plane. The $n^+$-fingers themselves are easily contacted with indium annealed at $350^\mathrm{o}$C.  We note that transport measurements in these structures cannot show parallel conduction since the fingers do not contact the modulation doping layer. The 2D carrier density can be tuned either by persistent photoconductivity or the sidegate. 4-point measurements were performed in a $^3\mathrm{He}$ cryostat, using lock-in techniques at an excitation frequency of 17~Hz.

The density and mobility can be characterized, and four contacts can be used to measure the longitudinal magnetoresistance $R_{xx}(B)$.  Figure \ref{nRXX} shows  $R_{xx}$ measured at 340 mK for an ungated sample, after illumination with a red LED.  The density deduced from Shubnikov-de-Haas (SdH) oscillations is $n=1.16 \times  10^{11}\: \mathrm{cm}^{-2}$. From the sample geometry, the square resistance $R_{\square}$ can be deduced giving the mobility $\mu =\frac{1}{R_{\square}ne} = 4.0 \times 10^6\: \mathrm{cm^2/Vs}$. At high magnetic fields up to 15 T, minima at fractional filling factors $ \nu$ = 1/3, 2/3, 2/5, 3/5, 3/7 and 4/7 can be clearly identified and attest the high quality of the sample.  One interesting and unexplained feature in these samples is the enhanced size of the fractional quantum Hall resistance maxima in comparison to the integer features.  The figure shows how the integer effect is magnified by a factor of $\times 10$ to be comparable in scale to the fractional effect.

\begin{figure}[t]
\center
\includegraphics[width=\linewidth ,keepaspectratio]{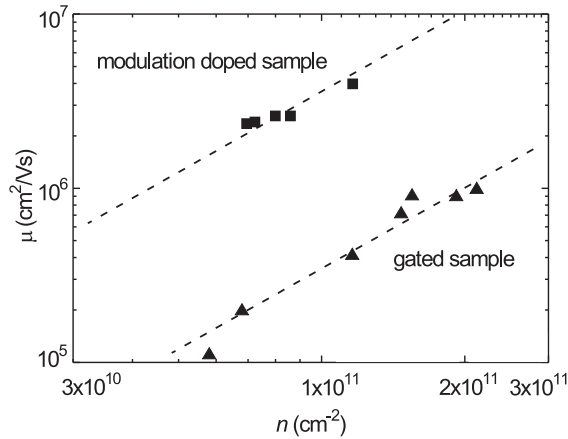}
\caption{Mobility $ \mu$ vs. electron density $n$ for a modulation-doped and a gated sample. Gated samples show an order of magnitude lower mobility at the same density. Dotted lines designate the power law $\mu = n^{1.5}$.}
\label{였sn}
\end{figure}  

Another advantage of the finger contacting technique is that gated CEO structures can be easily fabricated.  Figure \ref{였sn} compares the mobilities of a modulation doped sample and a gated sample at various carrier densities. In all samples the mobility increases with increasing density, showing the enhanced screening of the remote ionized dopants designated with empirical fit lines $\mu \sim n^{1.5}$.  The $n^+$-finger contacts give clear SdH oscillations at densities as low as $n = 6\times 10^{10}\: \mathrm{cm}^{-2}$.  The mobility of gated structures, however, is an order of magnitude lower than for the ungated structures at the same density. Gated structures typically have a reduced mobility relative to modulation doped structures, but the large difference observed here may arise due to heavy autocompensation typical to (110) $n$-doping.  

Second, we study growth and characterization of $p$-type CEO samples \cite{sample:pfinger}.  Silicon can be used as an acceptor in (110) GaAs if the arsenic pressure is lowered ($P_\mathrm{As} = 1.0 \times 10^{-5}$ mbar) and the sample temperature is increased ($T_\mathrm{pyr} = 670^\mathrm{o}$C) for the dopant layers, leaving the Si atoms to incorporate on the arsenic sites \cite{fischer:192106}.  The mutually perpendicular (110) and $(1\bar{1}0)$ crystal planes are suitable for $p$-type CEO, and serve as the primary growth and cleave regrowth directions, respectively.  The top layer of the substrate growth is a patterned $p^+$-layer above a $4\:\mu$m wide undoped GaAs layer.  The modulation doped layer in the cleave regrowth consists of $\rm 2\:nm$ of Si-doped $\rm Al_{0.3}Ga_{0.7}As$ separated by an 80 nm spacer from the cleave plane.   Even with highly doped $p$-fingers, the electrical contact to the 2D system at the cleave plane was highly resistive and improved by adding a $\rm 10\:nm$ thin GaAs well layer to the cleavage plane before the $\rm Al_{0.3}Ga_{0.7}As$ barrier.  Nonetheless, the contact resistance remained high, of order M$\Omega$.  

\begin{figure}[t]
\center
\includegraphics[width=\linewidth ,keepaspectratio]{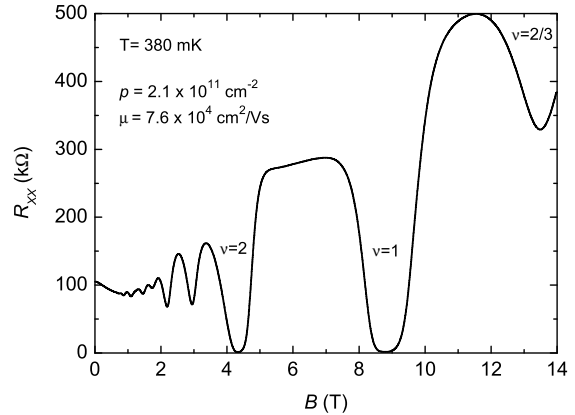}
\caption{Longitudinal resistance $\rm R_{xx}$ of a modulation doped 2D hole system fabricated with cleaved-edge overgrowth and modulated growth conditions. The minimum at $\nu $ = 2/3 indicates the onset of fractional quantum Hall effect.}
\label{pRXX}
\end{figure}

4-point measurements of $R_{xx}$ (Fig.~\ref{pRXX}) demonstrate the realization of a 2D hole system by means of CEO.  The pronounced minima at $\nu $ = 1 and 2 correspond to a hole density of $p = 2.1\times 10^{11}\: \mathrm{cm}^{-2}$. A measurement of $\rm R_{\square}$ yields the mobility $\rm \mu = 7.6 \times 10^4\: cm^2/Vs$. A fractional quantum Hall effect feature is visible at $\nu $ = 2/3. We note that in $p$-type samples, 4-point measurements are useful to observe the low magnetic field beating in the SdH-oscillations associated with the densities of the two spin split bands \cite{winkler:4245}. The mobilities achieved in CEO samples are about a factor of $3$ smaller than the highest reported values on (110) substrates doped with Si \cite{fischer:192106}.

In conclusion, a new sample design enables us to carry out four-point measurements on two-dimensional systems grown on in-situ cleaved facets. The CEO technique is used to fabricate both $n$- and $p$-type 2D systems while incorporating Si as the dopant in both cases. 4-point measurements of $R_{xx}$ at low temperatures show the fractional quantum Hall effect. Modulation-doped samples have a significantly better mobility than samples with an additional side gate. The realization of a 2D hole system on a cleaved facet is reported opening up further possibilities for p-type CEO.
\\
\\
Supported by the Deutsche Forschungsgemeinschaft via SFB 348 and Schwerpunktprogramm Quanten-Hall-Systeme and the Bundesministerium f\"ur Bildung und Forschung (BmBF) through project 01BM912.  M.G. would like to thank the A. v. Humboldt Foundation for support during this work.

%\begin{figure}[!ht]
%\center
%\includegraphics[width=\linewidth ,keepaspectratio]{n-finger3.eps}
%\caption{Longitudinal resistance $ R_{xx}$ of a modulation doped 2D electron system measured in a four-terminal geometry. Inset: lithographically defined $ n^+$-fingers conduct to the 2D system as current and voltage contacts. Minima at fractional filling factors $\nu$ prove the high quality of the sample.}
%\label{nRXX}
%\end{figure}

%\begin{figure}[t]
%\center
%\includegraphics[width=\linewidth ,keepaspectratio]{였sn.eps}
%\caption{Mobility $ \mu$ vs. electron density $n$ for a modulation-doped and a gated sample. Gated samples show an order of magnitude lower mobility at the same density. Dotted lines designate the power law $\mu = n^{1.5}$.}
%\label{였sn}
%\end{figure}  

%\begin{figure}[t]
%\center
%\includegraphics[width=\linewidth ,keepaspectratio]{p-finger.eps}
%\caption{Longitudinal resistance $\rm R_{xx}$ of a modulation doped 2D hole system fabricated with cleaved-edge overgrowth and modulated growth conditions. The minimum at $\nu $ = 2/3 indicates the onset of fractional quantum Hall effect.}
%\label{pRXX}
%\end{figure}

\end{document}